\documentclass[epj]{webofc}
\usepackage[varg]{txfonts}   

\usepackage{physics}
\usepackage[colorlinks=true]{hyperref}

\wocname{EPJ Web of Conferences}
\woctitle{ICNFP 2017}

\newcommand{\enquote}[1]{``#1''}

\newcommand{\NASixtyOne}{NA61\slash SHINE\xspace}

\newcommand{\NewMathSymbol}[2]{
  \newcommand{#1}{\ensuremath{#2}\xspace}
}
\NewMathSymbol{\y}{\mathrm{y}}
\NewMathSymbol{\pt}{p_T}
\NewMathSymbol{\mt}{m_T}
\NewMathSymbol{\sNN}{\sqrt{s_{NN}}}
\NewMathSymbol{\dE}{d_E}
\NewMathSymbol{\dEdx}{\text{dE/dx}}

\newcommand{\MeVVal}[1]{#1\,MeV\xspace}
\newcommand{\GeVVal}[1]{#1\,GeV\xspace}
\newcommand{\GeVValA}[1]{#1$A$\,GeV\xspace}
\newcommand{\sNNVal}[1]{\GeVVal{$\sNN = #1$}}
\newcommand{\topSPS}{\sNNVal{17.3}}

\let\myphi\phi
\renewcommand{\phi}{\ensuremath{\myphi}\xspace}

\newcommand{\withErrs}{with statistical (vertical lines) and systematic (bands) uncertainties}
\newcommand{\HorizontalLines}[1]{Horizontal lines give #1 bin sizes}

\makeatletter
\def\ignorecitefornumbering#1{%
     \begingroup
         \@fileswfalse
         #1
    \endgroup
}
\makeatother
\newcommand{\recite}[1]{Ref.~\cite{#1}}
\newcommand{\recites}[1]{Refs.~\cite{#1}}
\newcommand{\capcite}[1]{\ignorecitefornumbering{\cite{#1}}}
\newcommand{\caprecite}[1]{\ignorecitefornumbering{\recite{#1}}}
\newcommand{\caprecites}[1]{\ignorecitefornumbering{\recites{#1}}}

\newcommand{\Figref}[1]{Figure~\ref{#1}}
\newcommand{\figref}[1]{Fig.~\ref{#1}}
\newcommand{\figs}[1]{Figs.~\ref{#1}}
\newcommand{\equref}[1]{Eq.~(\ref{#1})}
\newcommand{\secref}[1]{section~\ref{#1}}

\newcommand{\figleft}{\emph{left}\xspace}
\newcommand{\figmiddle}{\emph{middle}\xspace}
\newcommand{\figright}{\emph{right}\xspace}

\newcommand{\pp}{p+p\xspace}
\newcommand{\PbPb}{Pb+Pb\xspace}
\newcommand{\BeBe}{Be+Be\xspace}
\newcommand{\AuAu}{Au+Au\xspace}
\newcommand{\ArSc}{Ar+Sc\xspace}
\newcommand{\SiSi}{\enquote{Si}+Si\xspace}
\newcommand{\CC}{\enquote{C}+C\xspace}

\NewMathSymbol{\totalYield}{\expval{\phi}}
\NewMathSymbol{\piYield}{\expval{\pi}}
\NewMathSymbol{\Kp}{K^+}
\NewMathSymbol{\Km}{K^-}

\newcommand{\NewSoftLibrary}[2]{
  \newcommand{#1}{\textsc{#2}\xspace}
}
\NewSoftLibrary{\Epos}{Epos}
\NewSoftLibrary{\CRMC}{Crmc}
\NewSoftLibrary{\UrQMD}{UrQMD}
\NewSoftLibrary{\Pythia}{Pythia}
\NewSoftLibrary{\HRG}{HRG}

\begin{document}

\selectlanguage{english}

\title{New baryonic and mesonic observables from \NASixtyOne}

\author{Antoni Marcinek\inst{1,2}\fnsep\thanks{\email{antoni.marcinek@gmail.com}} for the NA49 and \NASixtyOne collaborations
}

\institute{
  H. Niewodniczański Institute of Nuclear Physics, Polish Academy of Sciences, PL-31342 Kraków, Poland
  \and
  Jagiellonian University, Kraków, Poland
}

\abstract{%
  One of the main objectives of the \NASixtyOne experiment at the CERN SPS is to
  study properties of strongly interacting matter. This paper presents new
  results on observables relevant for this part of the \NASixtyOne programme.
  These include the first ever measurements of \phi meson production in \pp
  collisions at 40 and \GeVVal{80}, and most detailed ever experimental data at
  \GeVVal{158}. This contribution demonstrates the superior accuracy of the present dataset with
  respect to existing measurements. The comparison of \pp to \PbPb collisions
  shows a non-trivial system size dependence of the longitudinal evolution of
  hidden strangeness production, contrasting with that of other mesons.
  Furthermore, proton density fluctuations are investigated as a possible order
  parameter of the second order phase transition in the neighbourhood of the critical point
  (CP) of strongly interacting matter. An intermittency analysis is performed of
  the proton second scaled factorial moments in transverse momentum
  space. A previous analysis of this sort revealed significant power-law
  fluctuations for the \SiSi system at \GeVValA{158} measured by the NA49
  experiment. The fitted power-law exponent was consistent within errors with
  the theoretically expected critical value, a result suggesting a
  baryochemical potential in the vicinity of the CP of about \MeVVal{250}~\capcite{bib:NA49intermittency2015}.
  The analysis will now be extended to \NASixtyOne systems of similar size, \BeBe and \ArSc,
  at \GeVValA{150}. Finally, spectator-induced electromagnetic (EM) effects
  on charged meson production are being studied and bring information on the space-time position of the pion
  formation zone, which appears to be much closer to the spectator system for faster pions
  than for slower ones. On that basis, we demonstrate that the longitudinal evolution of the system at CERN
SPS energies may be interpreted as a pure consequence of local energy-momentum conservation.
}
\maketitle

\section{Introduction}
\begin{figure}[tb]
  \centering
  \includegraphics[width=\textwidth]{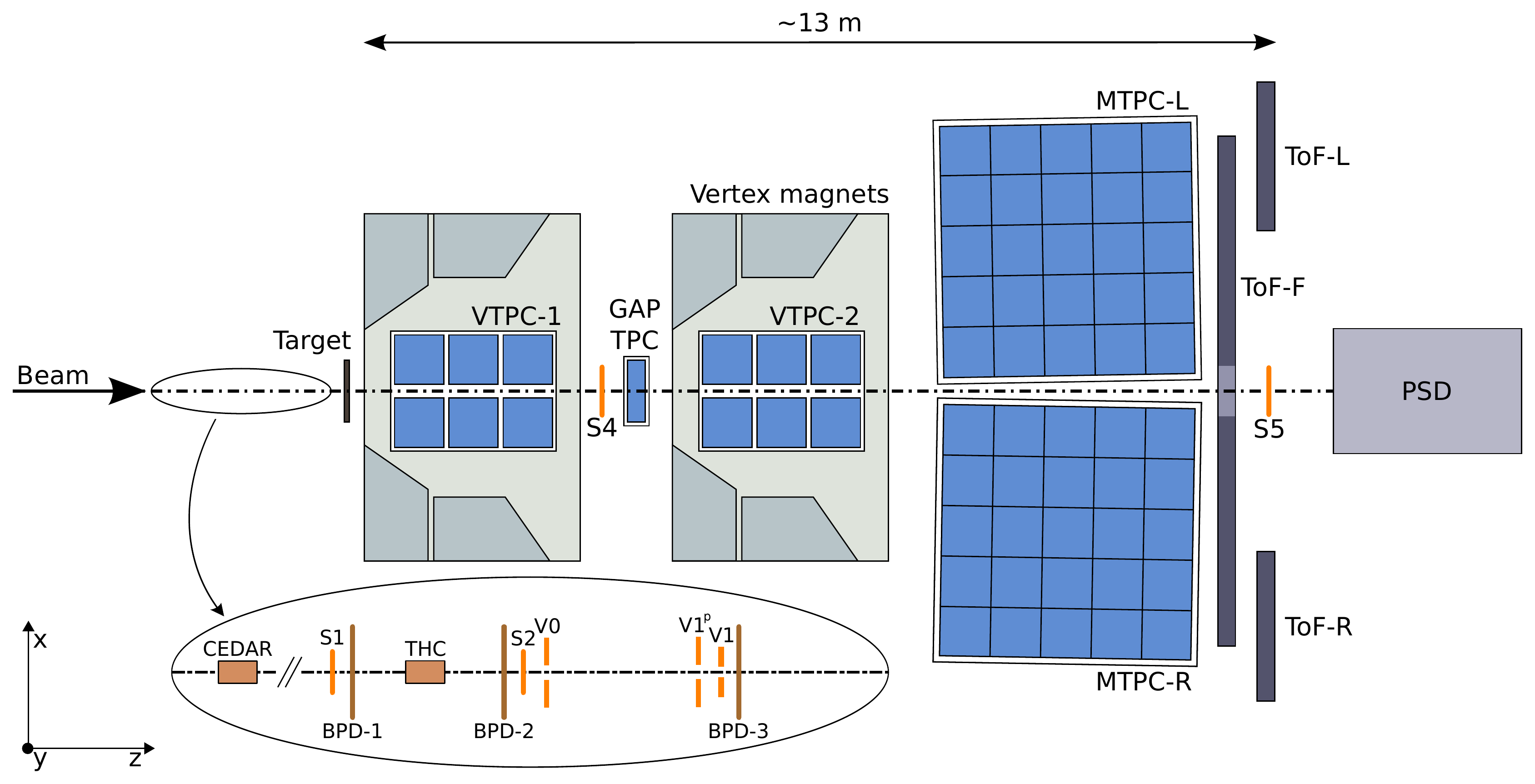}
  \caption[]{%
    Schematic layout of the \NASixtyOne detector
    system~\capcite{bib:NA61_facility} (horizontal cut in the beam plane, not
    to scale). Also outlined are the coordinate system used in the experiment
    and the configuration of beam detectors used with secondary proton beams for \pp
    interactions for which \phi meson production results are presented.
  }
  \label{fig:setup}
\end{figure}
\NASixtyOne~\cite{bib:NA61_facility} is a fixed target experiment at the CERN
SPS accelerator complex. Its strong interactions programme is a continuation
and extension of the NA49 experiment~\cite{bib:NA49_detector}, following NA49's
discovery of the onset of deconfinement in \PbPb collisions at \GeVValA{30}
beam momentum~\cite{bib:NA49Onset2002, bib:NA49Onset2008}. \NASixtyOne performs the
first two-dimensional scan of the phase diagram of strongly interacting matter
(SIM) by varying the momentum and size of the colliding nuclei, in search for
the critical point of SIM and to study properties of the onset of
deconfinement. Of interest to this programme are thus strangeness production
(\secref{s:phi}), fluctuations of baryon density (\secref{s:intermittency}) and
space-time properties of the medium created in heavy ion collisions
(\secref{s:EM}).
\par
The main parts of the \NASixtyOne detector system (\figref{fig:setup}) were
inherited from the NA49 experiment. Large volume Time Projection Chambers
(TPC), two of them immersed in vertical magnetic fields, allow precise
measurement of momenta of charged particles down to $\pt = 0$ and provide
particle identification (PID) via the \dEdx method. Time of Flight (ToF) walls
complement PID capabilities. The Projectile Spectator Detector (PSD), a hadron
calorimeter, measures forward-going energy allowing to estimate the centrality of
nuclear collisions.

\section{Hidden strangeness production in proton-proton collisions}\label{s:phi}
\begin{figure}[tb]
  \centering
  \includegraphics[width=\textwidth]{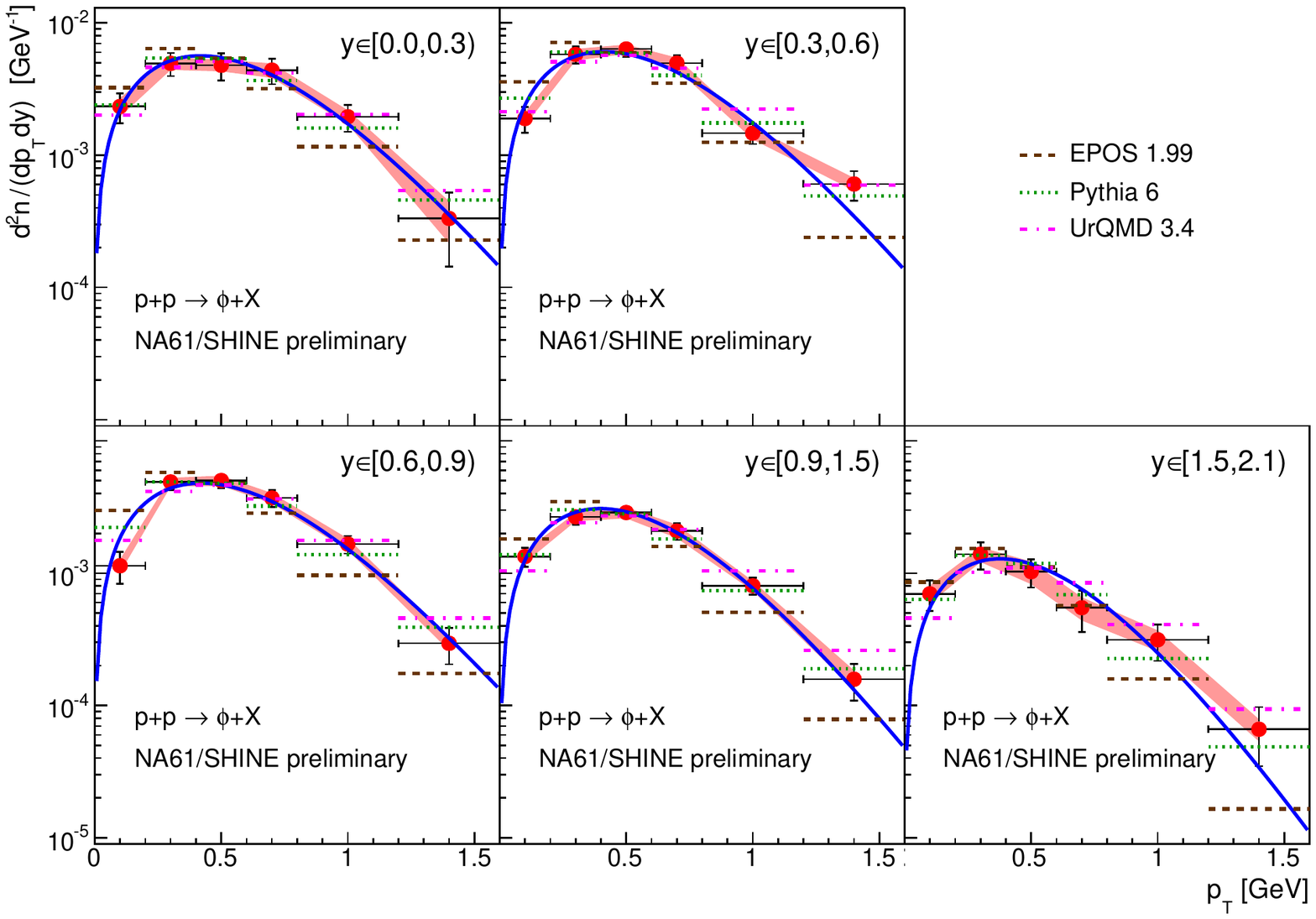}
  \caption{%
    Transverse momentum spectra in rapidity bins of \phi mesons in \pp
    collisions at \GeVVal{158} beam momentum (\sNNVal{17.3}) \withErrs.
    \HorizontalLines{\pt}. Regarding models and the fitted function see text.
  }
  \label{fig:pt158}
\end{figure}
\begin{figure}[tb]
  \centering
  \hspace*{-2ex}%
  \includegraphics[width=0.35\textwidth]{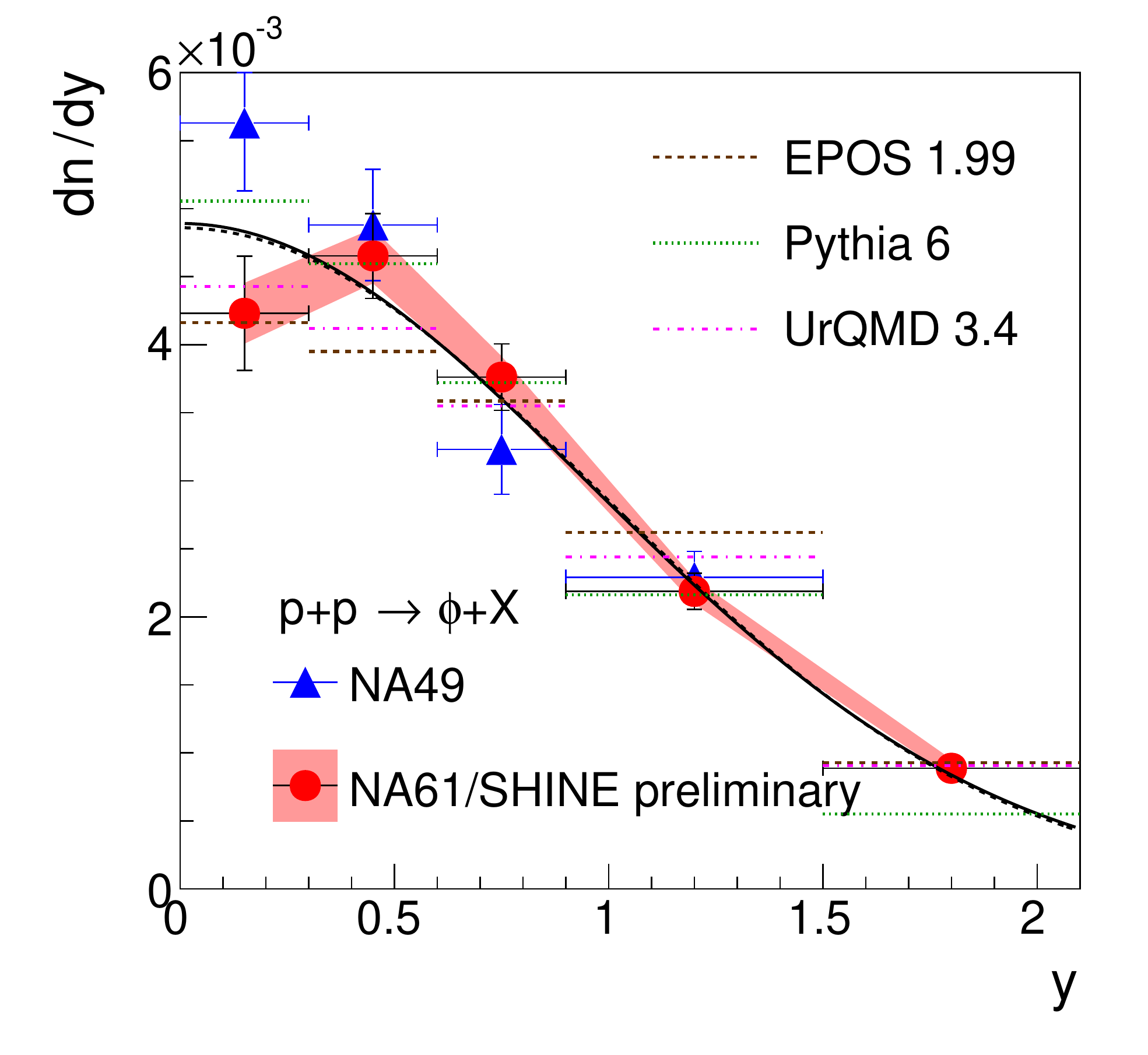}%
  \includegraphics[width=0.35\textwidth]{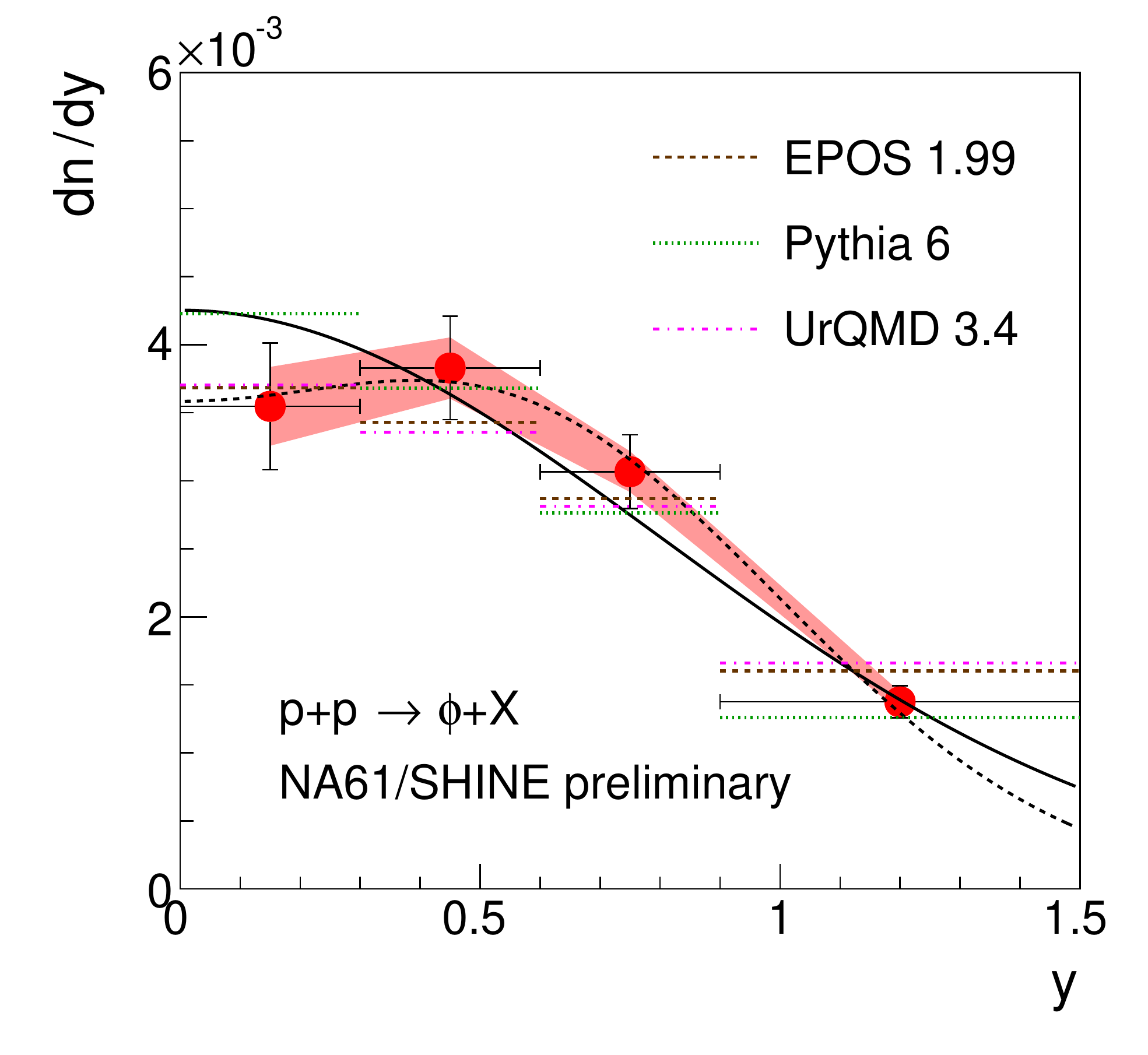}%
  \includegraphics[width=0.35\textwidth]{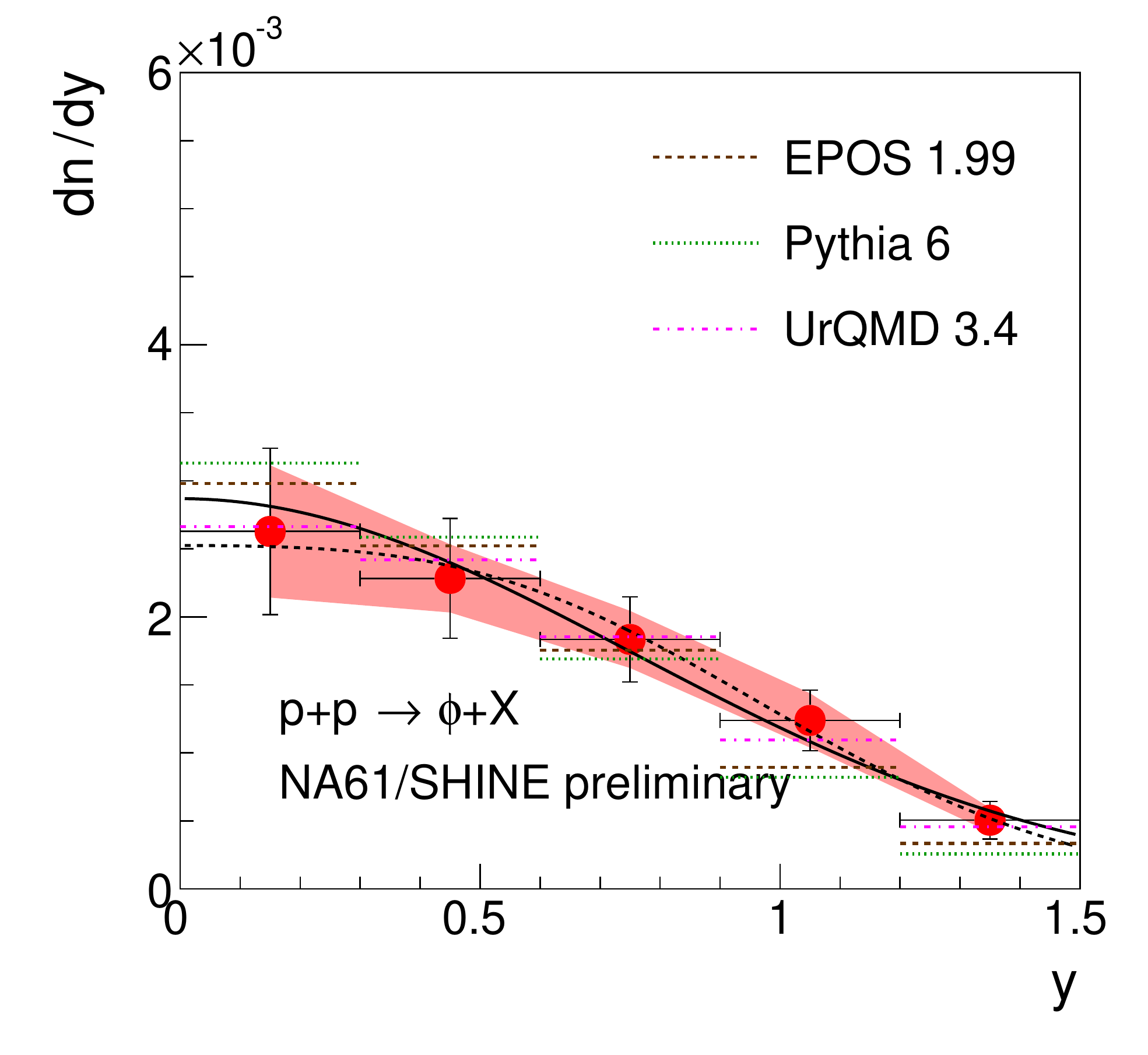}
  \caption[]{%
    Rapidity distributions of \phi mesons in \pp collisions at beam momenta of
    \GeVVal{158} (\figleft, \sNNVal{17.3}), \GeVVal{80} (\figmiddle,
    \sNNVal{12.3}) and \GeVVal{40} (\figright, \sNNVal{8.8}) \withErrs.
    \HorizontalLines{\y}. NA49 data points come from \caprecite{bib:NA49phi2000}.
    Regarding models and the fitted function see text.
  }
  \label{fig:rapidity}
\end{figure}
The \phi meson, consisting of $s$ and $\overline{s}$ valence quarks, is a particle with
hidden strangeness.  Motivation for measurement of its production in \pp
collisions is twofold.  First, it is useful in itself to constrain hadron
production models. Second, it may serve as reference for \PbPb measurements at
the same energies to identify strangeness-related phenomena specific to heavy ion
collisions.
\par
In this contribution, differential multiplicities of \phi mesons in \pp interactions at
40, 80 and \GeVVal{158} beam momenta, as a function of rapidity~\y and
transverse momentum~\pt, are obtained from invariant mass spectra in the $\phi
\to \Kp\Km$ decay channel. In \figref{fig:pt158} one can see \pt spectra in
rapidity bins for the \GeVVal{158} data set. Similar double-differential spectra
are also measured for \GeVVal{80}, while due to low track statistics, only
single-differential analysis, separately in \y and in \pt, was feasible for the
\GeVVal{40} data. These are the first ever differential measurements of \phi
production at 40 and \GeVVal{80} and the first double-differential measurements
at \GeVVal{158}.
\par
Transverse momentum spectra are compared to three models:
\Epos~1.99~\cite{bib:EPOS2006, bib:EPOS2009} from the \CRMC~1.6.0
package~\cite{bib:CRMC}, \Pythia~6.4.28~\cite{bib:PythiaManual} and
\UrQMD~3.4~\cite{bib:UrQMD1998, bib:UrQMD1999}. Model predictions in \figref{fig:pt158}
are normalized to the integral of the data in each rapidity bin to focus only on
shape comparison. One sees that \Pythia describes well the shape, while the spectra
from \UrQMD are too hard and those from \Epos too soft.
\par
Having the double differential spectra it is possible to calculate the single
differential spectra of rapidity integrated over \pt by summation of the
measured \pt region and extrapolation to large \pt. The latter is done using a
thermally motivated fit function $\pt e^{\mt/T}$ (thick curves in
\figref{fig:pt158}). The unmeasured tail contribution is smaller than 1\% for
all \y bins in \figref{fig:pt158}.
\par
The resulting rapidity distributions are shown in \figref{fig:rapidity} for all
studied energies. Again the shape comparison with the three models is
performed using the same normalization scheme as for \pt spectra. In this case
both \Epos and \UrQMD predict shapes comparable to that of the data, while \Pythia
produces a distribution which is too narrow. For the largest energy, single differential
measurements in narrower rapidity range, are available from
NA49~\cite{bib:NA49phi2000}. It is clear that the two measurements are
consistent.
\begin{figure}[tb]
  \centering
  \sidecaption
  \includegraphics[width=0.5\textwidth,page=1]{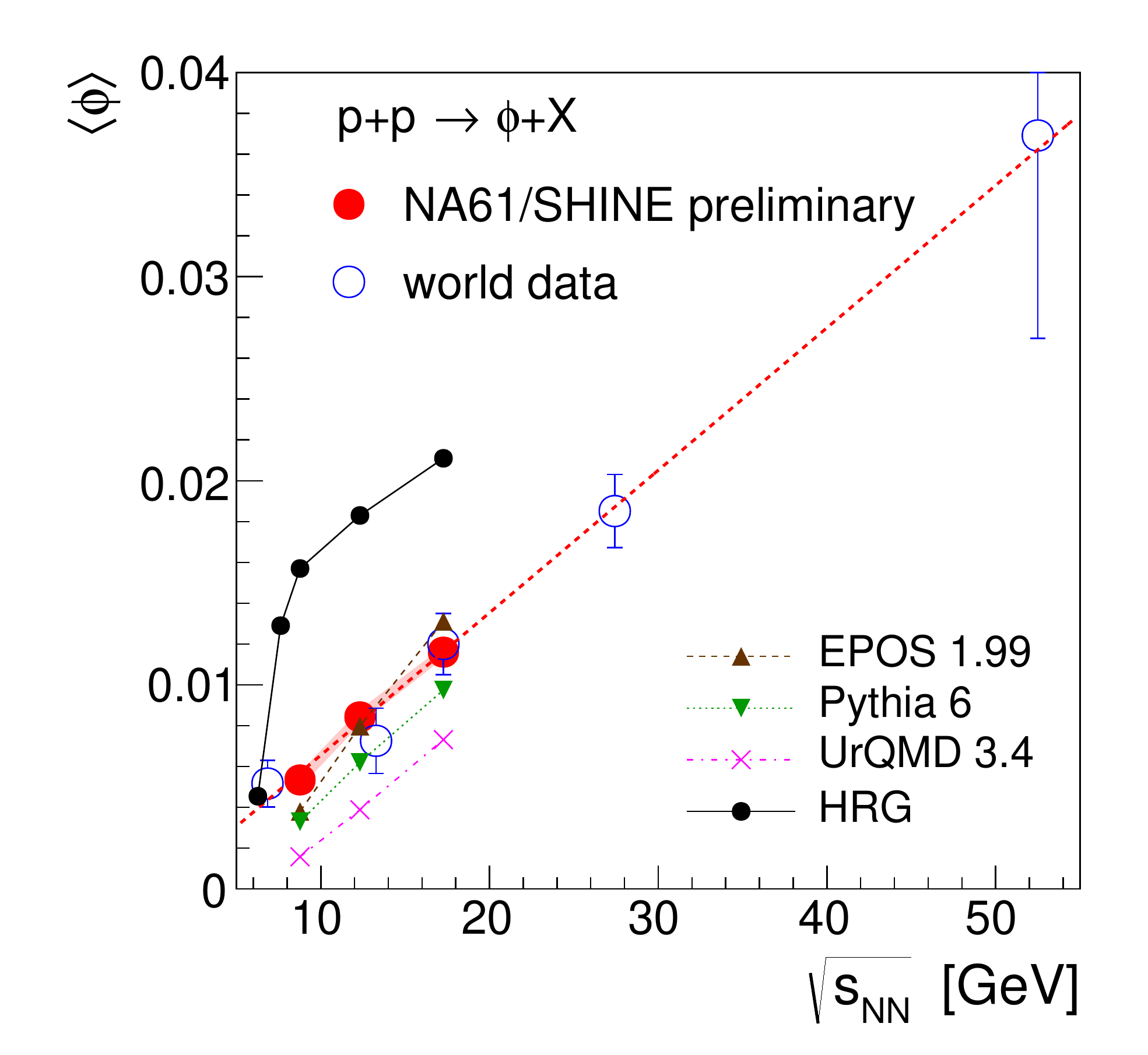}%
  \caption[]{%
    Energy dependence of total yields of \phi~mesons in \pp collisions. World
    data come from \caprecites{bib:Blobel_pp24_phi_1975,
    bib:ACCMOR_hadrons63_93_phi_1981, bib:Drijard_ppS53_phi_1981,
    bib:AguilarBenitez_pp400_phi_1991, bib:NA49phi2000}. Total uncertainties
    are shown; for preliminary \NASixtyOne results total uncertainties are
    smaller than markers. Red dashed line is to guide the eye. Regarding
    models see text.
  }
  \label{fig:yields}
\end{figure}
\par
In order to calculate total \phi multiplicities, distributions are fitted with
Gaussian functions\footnote{%
It should be noted that for \phi meson rapidity distributions
a double Gaussian (dashed curves in \figref{fig:rapidity}) does not provide any
advantage over a single Gaussian function contrary to the case of charged hadrons in
\pp~\capcite{bib:NA61_pp_spectra_2017}.%
} (solid curves in \figref{fig:rapidity}). Again summation of the measured
spectrum is done and the unmeasured tail contribution from the fit is added (3\% to
7\% depending on energy). Then the result is doubled to account for the
backward hemisphere from collision symmetry. The energy dependence of the obtained
total \phi yields is shown in \figref{fig:yields} and compared to world
data~\cite{bib:Blobel_pp24_phi_1975, bib:ACCMOR_hadrons63_93_phi_1981,
bib:Drijard_ppS53_phi_1981, bib:AguilarBenitez_pp400_phi_1991,
bib:NA49phi2000}. Clearly \NASixtyOne results are consistent with the world
data, but are much more accurate. The figure also shows a comparison of \phi
yields with models, including the hadron resonance gas
(\HRG~\cite{bib:Vovchenko2016}) statistical model. It is apparent that while
\Epos describes the data reasonably well (although the rise with
collision energy is too fast), all other models fail with \UrQMD underestimating and \HRG
overestimating the yield by about a factor of 2.
\begin{figure}[tb]
  \centering
  \includegraphics[width=0.5\textwidth,page=1]{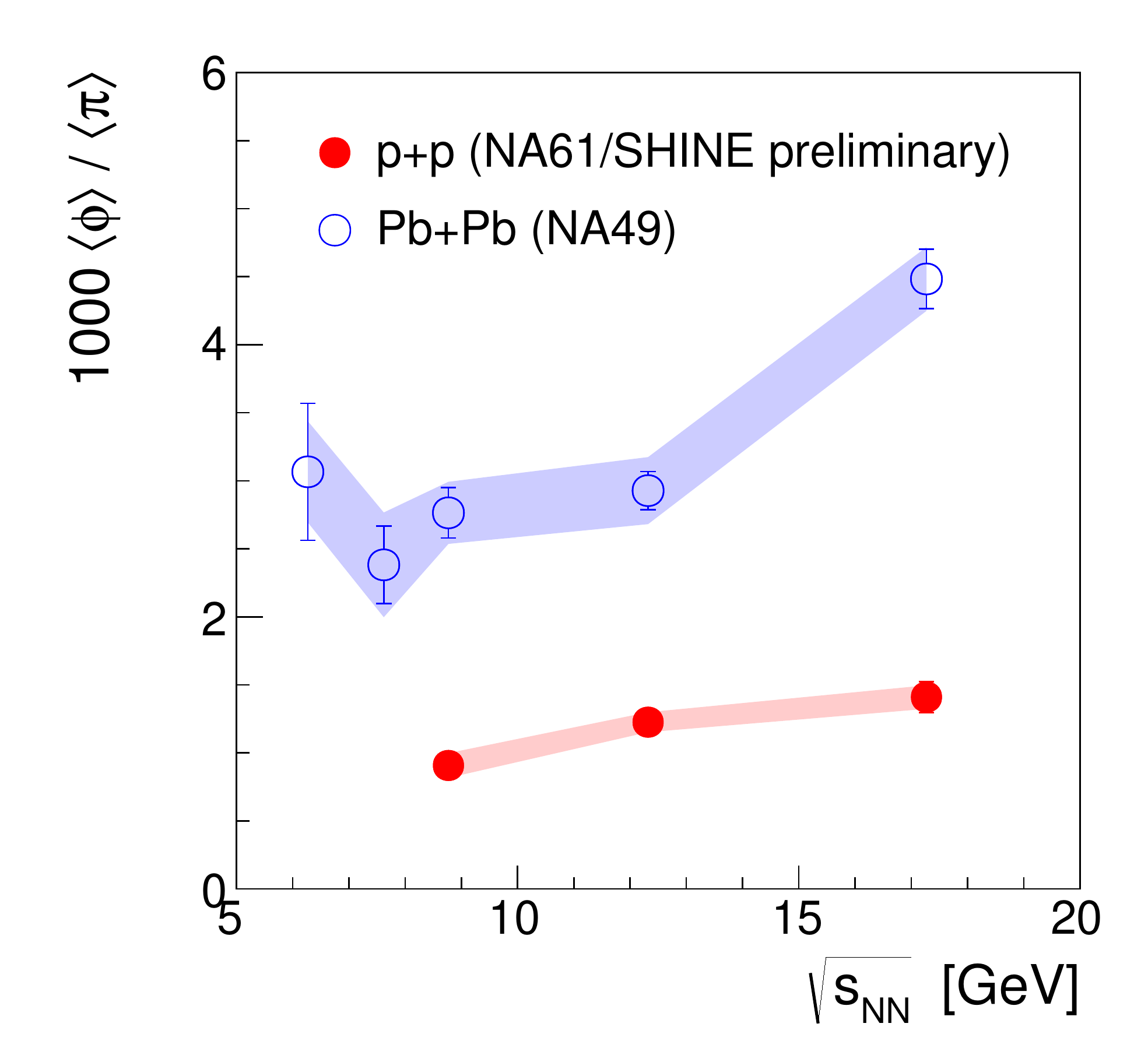}%
  \includegraphics[width=0.5\textwidth,page=2]{figs/YieldRatios.pdf}
  \caption[]{%
    Energy dependence of (\figleft) ratios of total yields of \phi mesons to
    mean total yields for pions \equref{eq:piYield} in \pp and \PbPb,
    (\figright) double ratios (see text), \withErrs.  Full circles correspond
    to results of this analysis, \PbPb data come from
    NA49~\capcite{bib:NA49phi2008, bib:NA49Onset2002, bib:NA49Onset2008}, while
    \pp kaon and pion data are taken from \caprecite{bib:NA61_pp_spectra_2017}.
  }
  \label{fig:YieldRatios}
\end{figure}
\par
\Figref{fig:YieldRatios} demonstrates the enhancement of strange meson
production in \PbPb collisions compared to \pp reactions. Data on \phi in \PbPb and
other mesons in both systems come from \recites{bib:NA49phi2008,
bib:NA49Onset2002, bib:NA49Onset2008, bib:NA61_pp_spectra_2017}. Multiplicities
of strange mesons are divided by those of pions to take out the
\enquote{trivial} effect of the size of the colliding system. In case of the \phi, the mean
total yield of pions is used which was calculated as \recite{bib:NA49phi2008}:
\begin{equation}
  \label{eq:piYield}
  \textstyle
  \piYield = \dfrac{3}{2} \pqty{\expval{\pi^+} + \expval{\pi^-}} \,.
\end{equation}
It is evident from \figref{fig:YieldRatios}~(\figleft), that the $\phi/\pi$ ratio
increases with energy for both systems. The enhancement from p+p to central Pb+Pb collisions
is about a factor 3.
\par
Dividing the \PbPb by the \pp ratios one gets the double ratio:
\NewMathSymbol{\yieldsRatio}{\totalYield / \piYield}
\NewMathSymbol{\yieldsRatioQty}{\pqty{\yieldsRatio}}
\begin{equation}
  \operatorname{double~ratio}\yieldsRatioQty =
    \frac{\yieldsRatioQty_\text{\PbPb}}{\yieldsRatioQty_\text{\pp}} \,,
\end{equation}
which is shown by the full circles in \figref{fig:YieldRatios}~(\figright).
The analogous ratio for kaons is also plotted in \figref{fig:YieldRatios}~(\figright). 
It is apparent that the enhancement of \phi production is comparable to that of \Kp,
while the enhancement for \Km is systematically smaller.
Furthermore the enhancement of \phi production is about the square of the \Km 
enhancement, consistent with the hypothesis of strangeness enhancement in a 
parton phase of the collision.
\begin{figure}[tb]
  \centering
  \sidecaption
  \includegraphics[width=0.5\textwidth,page=3]{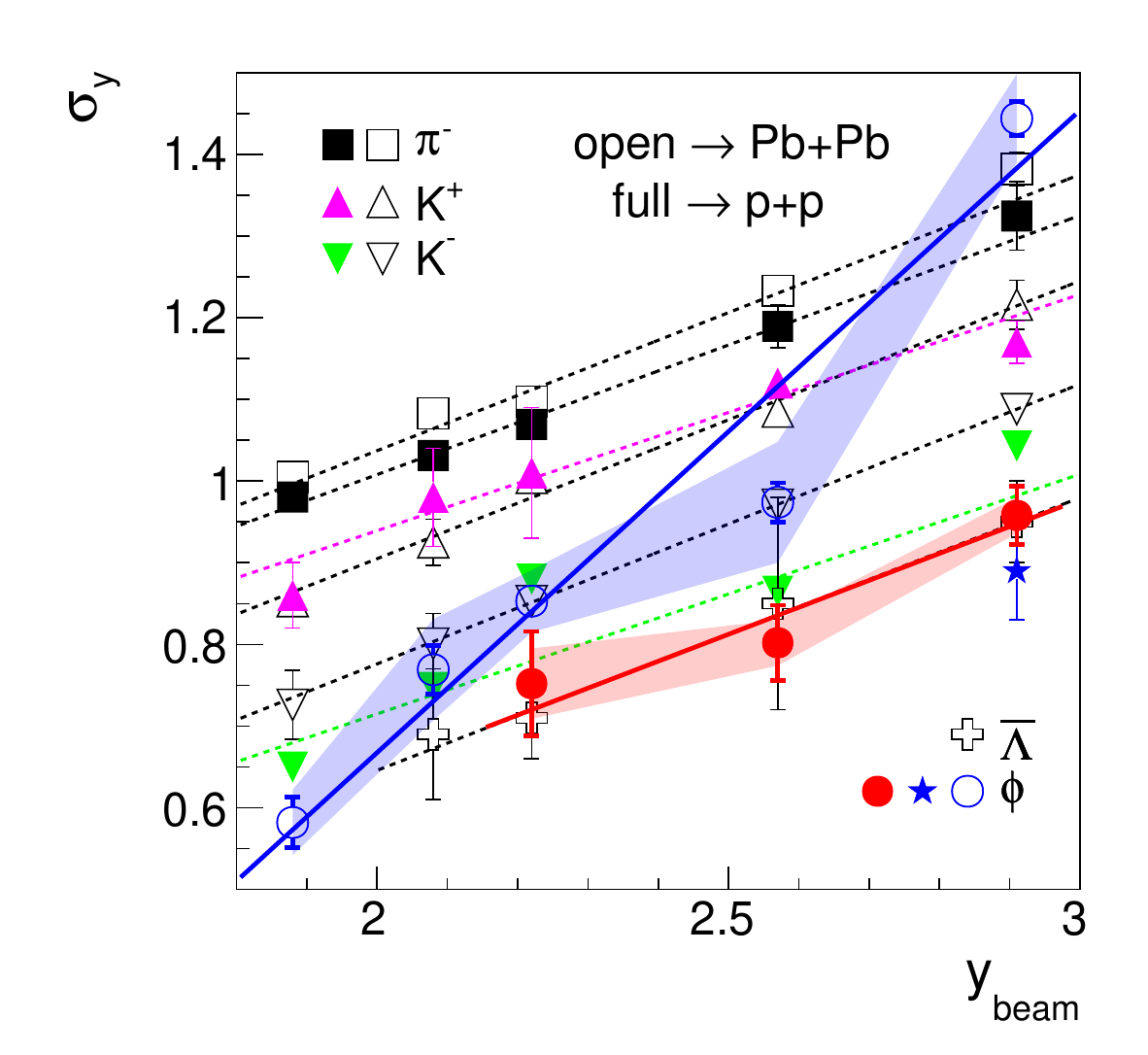}%
  \caption[]{%
    Widths of rapidity distributions of various particles in \pp (full symbols)
    and central \PbPb (open symbols) collisions as a function of beam rapidity,
    \withErrs. Full circles are results of this analysis, the star is a NA49
    measurement in \pp~\capcite{bib:NA49phi2000}, other \pp points come from
    \NASixtyOne~\capcite{bib:NA61_pp_pion_2014, bib:NA61_pp_spectra_2017}.
    \PbPb data are from NA49~\capcite{bib:NA49phi2008, bib:NA49Onset2002,
    bib:NA49Onset2008, bib:NA49Lambdas2004}. Lines are fitted to points to
    guide the eye.
  }
  \label{fig:sigma}
\end{figure}
\par
\NewMathSymbol{\sigmaY}{\sigma_\y}
\NewMathSymbol{\ybeam}{\y_\text{beam}}
Finally, \figref{fig:sigma} shows the widths \sigmaY of rapidity distributions of
\phi mesons and various other particles in \pp and central \PbPb
collisions~\cite{bib:NA49phi2000, bib:NA61_pp_pion_2014,
bib:NA61_pp_spectra_2017, bib:NA49phi2008, bib:NA49Onset2002,
bib:NA49Onset2008, bib:NA49Lambdas2004} as a function of beam rapidity \ybeam
in the centre-of-mass frame. Widths are derived from single or double Gaussian
fits depending on the particle species. It is striking that all particles in
both colliding systems, \emph{except \phi in \PbPb} follow the same trend:
\sigmaY is proportional to \ybeam in the studied range, with the same slope
of increase with energy.
While the effect was already known from \recite{bib:NA49phi2008}, the new
\NASixtyOne results emphasize the peculiarity of the system size dependence of the
longitudinal evolution of \phi production, contrasting with that of other
mesons.
\par
It should be noted, that the behaviour of \sigmaY for \phi mesons in \PbPb collisions is
qualitatively consistent with rescattering of kaons from \phi mesons decaying
inside the fireball~\cite{bib:JohnsonRescattering, bib:NA49phi2008}. Such kaons
no longer contribute to the signal peak in the invariant mass spectrum. It is
more likely to take place for slow kaons (i.e.\ those coming from low rapidity
\phi mesons), which travel longer through the fireball. Therefore, the \phi
rapidity spectrum gets depleted and this loss is the highest at midrapidity and
decreases with increasing \y values. So the spectrum becomes wider due to the
rescattering. The effect should increase with the collision energy, because the
higher the energy, the larger and denser the fireball. Moreover, it is
natural to assume that the effect would be much stronger in a
quark gluon plasma, than in a purely hadronic medium. This shows the importance
 of investigating the system size dependence of the effect for systems between \pp and
\PbPb, for which data are collected by \NASixtyOne.

\section{Search for the QCD critical point via intermittency analysis}\label{s:intermittency}
\begin{figure}[tb]
  \centering
  \begin{minipage}[b]{0.8\textwidth}
    \includegraphics[width=\textwidth]{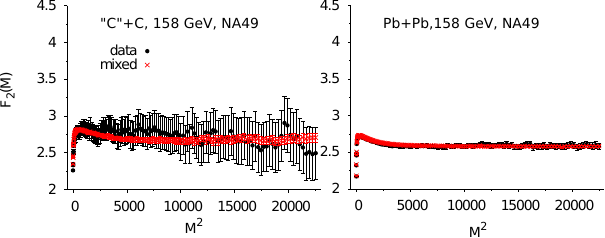}
    \begin{minipage}[b]{0.5\textwidth}
      \includegraphics[width=\textwidth]{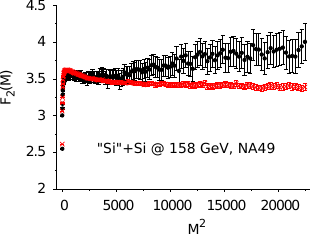}
      \includegraphics[width=\textwidth]{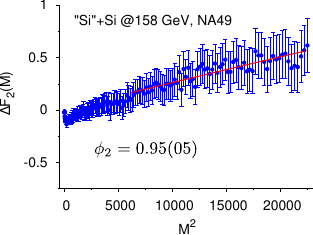}
    \end{minipage}%
    \begin{minipage}[b]{0.48\textwidth}
      \hspace{2ex}\includegraphics[width=\textwidth,clip,trim=0 3mm 7mm 2mm]{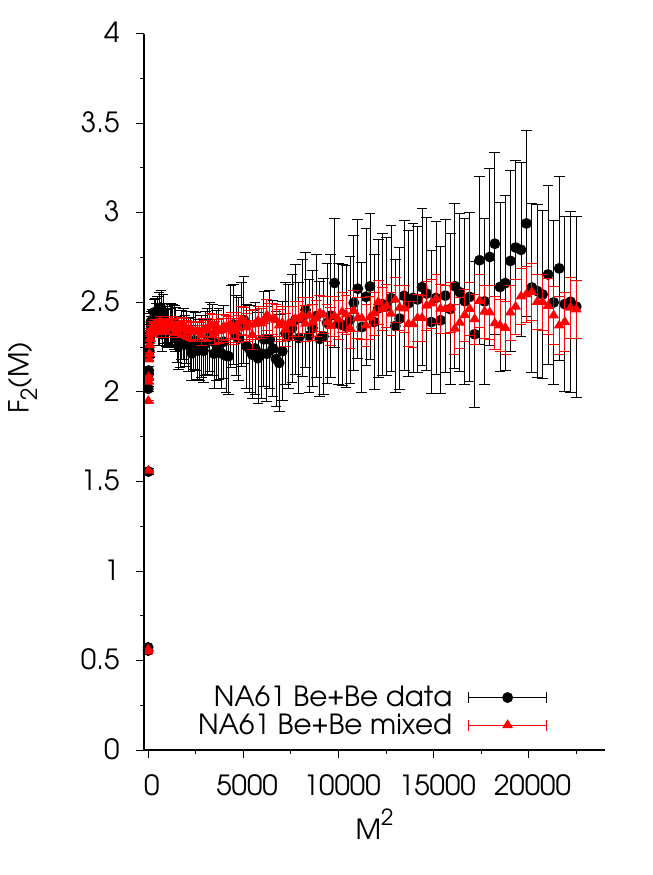}
    \end{minipage}
  \end{minipage}
  \caption[]{%
    Second scaled factorial moments in transverse momentum space for protons in
    midrapidity for various colliding systems at top CERN SPS energy, measured
    by NA49~\capcite{bib:NA49intermittency2015} and \NASixtyOne.
  }
  \label{fig:intermittency}
\end{figure}
In the vicinity of the critical point of strongly interacting matter increased
fluctuations in the system are expected in analogy to critical
opalescence in liquid-vapour phase transitions. Therefore the procedure of the
search for the CP is to scan the phase diagram of SIM by varying energy and size of
colliding nuclei and look for a maximum in fluctuation observables.
\par
Among various fluctuation measures, this contribution focuses on local, power-law
fluctuations of baryon density. These are studied by calculating second scaled
factorial moments $F_2$ in transverse momentum space ($p_y$ vs.\ $p_x$) of
protons in midrapidity:
\newcommand{\expnn}[1]{\expval{n_{#1} \pqty{n_{#1} - 1}}}
\newcommand{\dsum}{\displaystyle\sum}
\begin{equation}
  F_2(M) \equiv \frac{\expval{\dsum_m n_m \pqty{n_m - 1}}}{\expval{\dsum_m n_m^2}} \,,
  \label{eq:F2}
\end{equation}
where the $(p_x, p_y)$ space is divided into $M$ bins, $n_m$ is the number of
protons in bin number $m$ and $\expval{\ldots}$ denotes averaging over events.
For a critical system, $F_2$ is expected to scale according to a specific power
law with the bin size or equivalently number of bins~$M$. This scaling is
called \emph{intermittency} and the corresponding exponent is called the
\emph{intermittency index} $\phi_2$~\cite{bib:Diakonos_CPOD2006}. However, it
is also expected that in addition to protons stemming from a critical system, there is
some non-critical background. The latter is estimated using event mixing. So
finally the correlator
\begin{equation}
  \Delta F_2(M) = F_2^{\text{data}}(M) - F_2^{\text{mix}}(M)
  \label{eq:DF2}
\end{equation}
should scale for $M \gg 1$ proportionally to $(M^2)^{\varphi_2}$ with
intermittency index $\varphi_2 = 5/6$~\cite{bib:Antoniou_2006}.
\par
\Figref{fig:intermittency} shows $F_2(M)$ as measured by NA49
for several systems at \GeVValA{158}~\cite{bib:NA49intermittency2015}. It is
seen that data and mixed events plots overlap for the \CC and \PbPb systems\footnote{%
Here \enquote{C} (\enquote{Si}) actually are a mixture of nuclei with Z=6,7 (13,14,15).}, so
no intermittency is observed there. In the \SiSi system, on the other hand, there
is a clear separation of data and mixed events, so $\Delta F_2(M)$ can be
calculated and fitted. The intermittency index $\varphi_2 =
0.96^{+0.38}_{-0.25} (\text{stat.}) \pm 0.16 (\text{syst.})$ thus obtained
is consistent, within uncertainties,\footnote{%
Statistical uncertainty was estimated from the bootstrap
method~\cite{bib:NA49intermittency2015}.} with the theoretical predictions for
a critical system.  \Figref{fig:intermittency}  also shows $F_2(M)$ for the \BeBe system measured by
\NASixtyOne at the top SPS energy. No evidence for intermittency is observed
there, consistent with NA49 results, as the \BeBe system is lighter than
\CC and therefore yet further away from the \SiSi system. Finally, the analysis for the heavier
\ArSc system in \NASixtyOne is ongoing \cite{bib:statReport2017}.

\section{Electromagnetic effects in pion emission}\label{s:EM}
\NewMathSymbol{\Excen}{E^*_\text{s}}
\NewMathSymbol{\ys}{\y_\text{s}}
\begin{figure}[tb]
  \centering
  \includegraphics[width=0.8\textwidth]{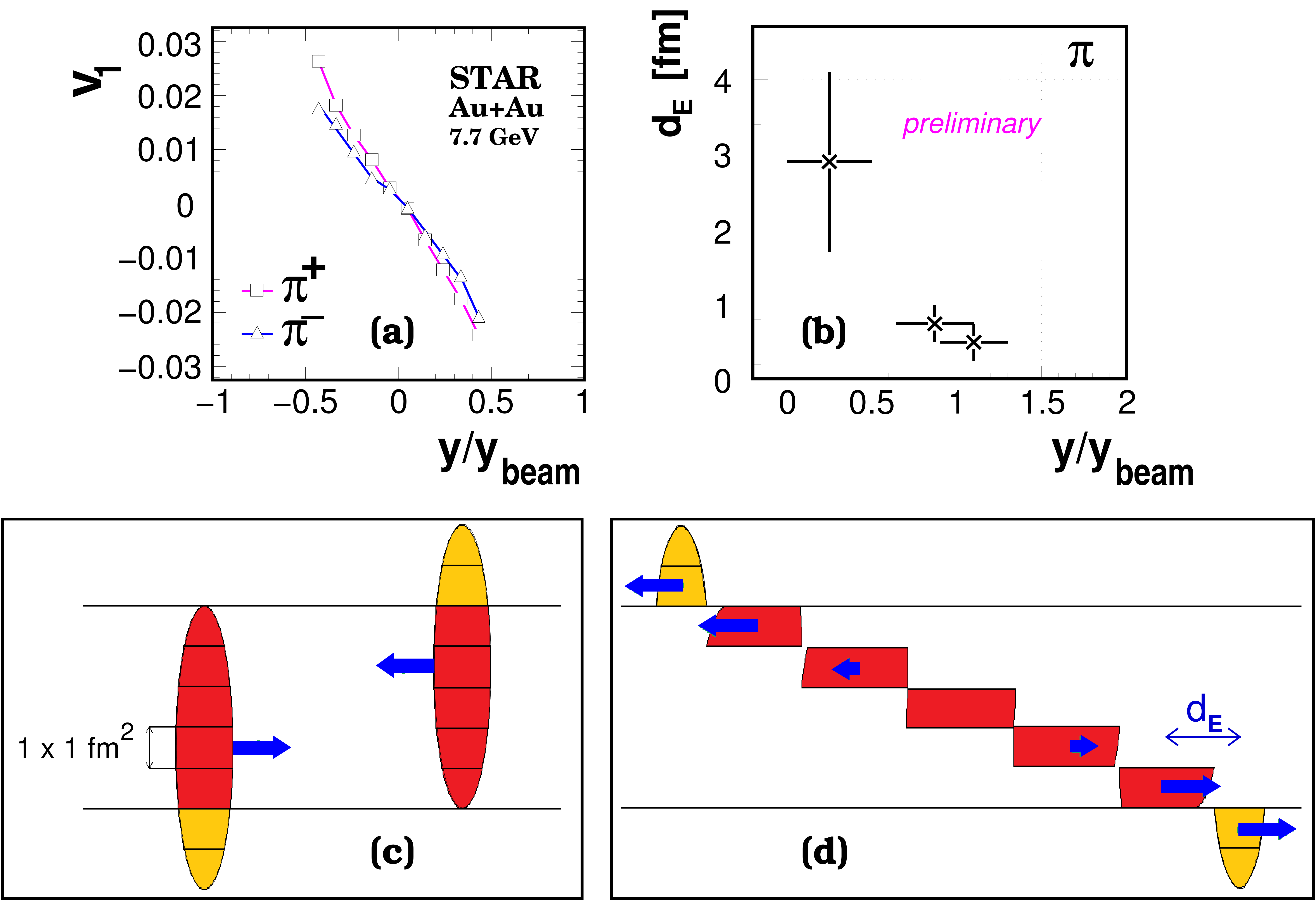}
  \caption[]{%
    (a)~Directed flow for $\pi^+$ and $\pi^-$ in Au+Au
    collisions~\capcite{bib:Rybicki_2015}; original data from
    \caprecite{bib:STAR_2014}. (b)~Dependence of \dE on pion rapidity for Au+Au
    and Pb+Pb collisions at RHIC BES/CERN SPS energies~\capcite{bib:Rybicki_2016}.
    (c)~\enquote{Bricks} of matter considered in the fire streak model of
    \caprecite{bib:Szczurek_2017} before the collision.  (d)~Fire streaks formed
    after the collision, also redrawn from \caprecite{bib:Szczurek_2017}. Thick
    arrows indicate velocity vectors.
  }
  \label{fig:em1}
\end{figure}
\begin{figure}[tb]
  \centering
  \includegraphics[width=\textwidth]{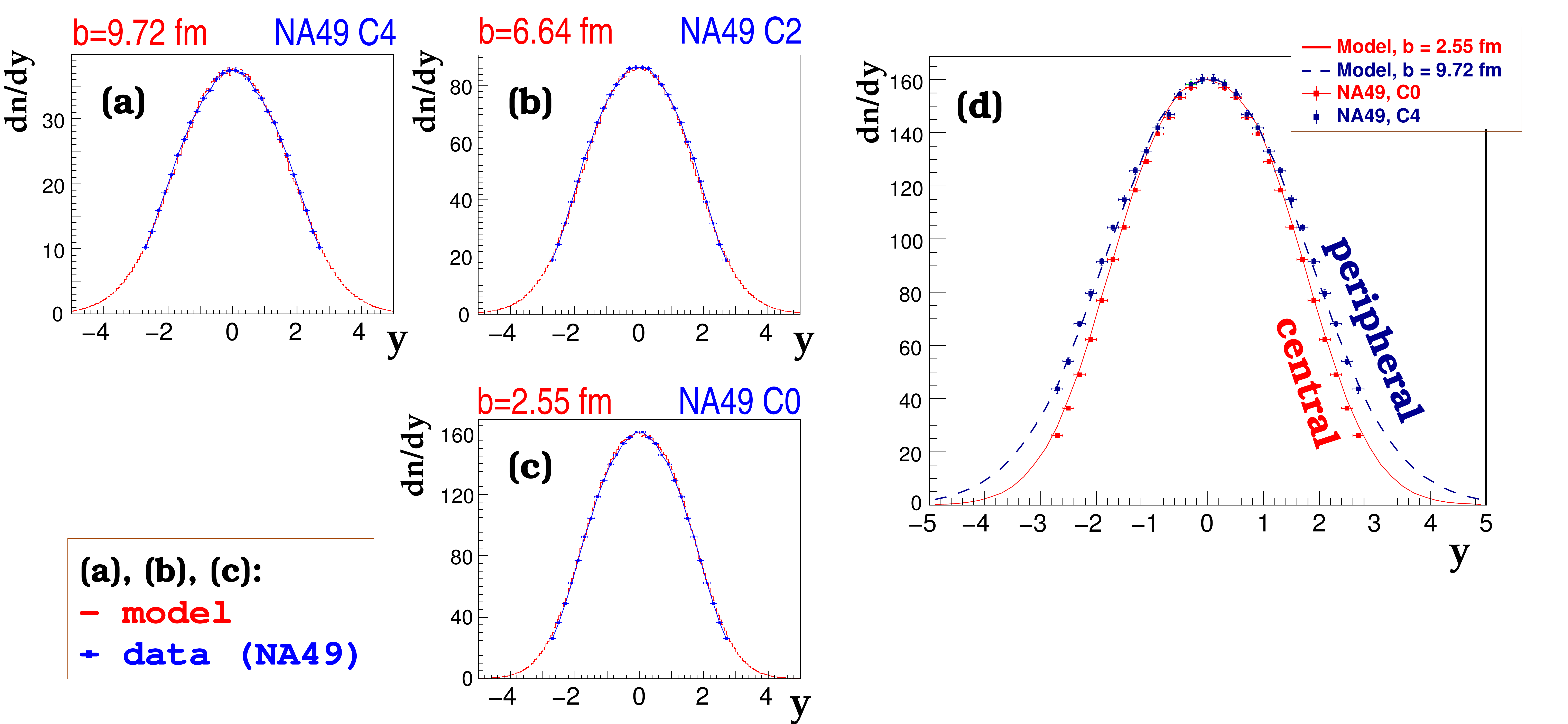}
  \caption[]{%
    (a)--(c)~Rapidity distributions of negatively charged pions measured by the NA49
    experiment in three centrality classes, from the most peripheral (C4) to the
    most central (C0) \PbPb collisions at \topSPS, compared to the fire streak
    model~\capcite{bib:Szczurek_2017}; experimental data are taken from
    \caprecite{bib:NA49_2012}. (d)~Shape comparison of rapidity spectra for two
    extreme centrality classes together with the fire streak model. The
    spectrum for peripheral collisions is renormalized at its peak to central
    collisions.
  }
  \label{fig:em2}
\end{figure}
Charged spectators in non-central heavy ion collisions generate electromagnetic
fields, which modify trajectories of final state charged particles (most of
them are pions). Oppositely charged particles are affected oppositely,
leading to charge asymmetries in distributions of produced particles. One example of such
asymmetries is charge splitting of directed flow~$v_1$ visible in
\figref{fig:em1}(a) for \AuAu collisions measured by the STAR
experiment~\cite{bib:STAR_2014}. It was explained by the spectator-induced
electromagnetic (EM) effects assuming that the distance \dE between the pion
formation zone and the spectator was of the order of
3\,fm~\cite{bib:Rybicki_2015}.
\par
Taking together results of three similar analyses of EM effects in three
different data sets on Au+Au and Pb+Pb
collisions~\cite{bib:STAR_2014,bib:Rybicki_2009,bib:wa98}, the dependence of \dE on
pion rapidity, shown in \figref{fig:em1}(b), was
obtained~\cite{bib:Rybicki_2016}. It is visible, that faster pions are produced
closer to the spectator system.
\par
The latter observation was explained in \recite{bib:Szczurek_2017} in a new,
independent realization of the fire streak model~\cite{bib:Myers_1978},
sketched in \figs{fig:em1}(c) and (d). In the collision centre-of-mass frame,
the two incoming nuclei are represented as two continuous, Lorentz-contracted
3D mass distributions defined by the known nuclear density
profiles~\cite{bib:trzcinska1}. They are then divided into
\enquote{bricks} in the transverse plane of the collision. Each brick collides
independently with the one from the opposite nucleus, forming a \enquote{fire
streak}. For each pair, local energy-momentum conservation is assumed, which
yields the excitation energy~\Excen and rapidity~\ys for each fire
streak.\footnote{%
For the actual non-trivial distribution of \Excen and \ys in the transverse
space as a function of collision centrality, and for the detailed discussion of
normalisation issues in the comparison of the model to the experimental data
discussed later in the text, the reader is referred to
\recite{bib:Szczurek_2017}.} The final ingredient of the model is that each
fire streak fragments independently into $n$ pions, according to the
function~\cite{bib:Szczurek_2017}:
\begin{equation}
  \frac{\mathrm{d}n}{\mathrm{d}\y}
  \sim A \cdot (\Excen - m_\text{s})
  \cdot \exp\left(- \frac{[(\y - \ys)^2 + \epsilon^2]^\frac{r}{2}}{r \sigma_\y^r}\right)
  \label{eq:fragment}
\end{equation}
of pion rapidity~\y with free parameters $A$, $r$, $\sigma_\y$
\emph{independent of the fire streak and collision centrality}, $\epsilon =
0.01$, and other parameters describing properties of the fire streak stemming
directly from the conservation rule. From the distribution of fire streak
velocities, depicted as thick arrows in \figref{fig:em1}(d), and
formula~(\ref{eq:fragment}) one finds that on average, pions produced
closer to the spectator system are moving faster.
\par
Summing formula~(\ref{eq:fragment}) over all fire streaks, one obtains the full
rapidity spectrum of pions. The fit of the model to data from
NA49~\cite{bib:NA49_2012} for different centrality classes is shown in
\figref{fig:em2}. It is apparent that the model describes well the dependence
on centrality of both the yields of pions and shapes of pion rapidity spectra.
This means that the longitudinal evolution of the system at CERN SPS energies
can be interpreted as a \emph{pure consequence of local energy-momentum
conservation}.
\par
Finally, it should be noted that apart from three measurements leading to
the results in \figref{fig:em1}(b), an analysis of EM effects for \ArSc collisions in
the \NASixtyOne experiment is ongoing~\cite{bib:statReport2017}.

\section{Summary}
New results on observables relevant for the strong interactions programme of
the \NASixtyOne experiment were presented. They include the first ever
differential measurements of \phi meson production in \pp collisions at 40 and
\GeVVal{80} beam momentum and the first double-differential measurements at
\GeVVal{158}. These results were compared to world data on \pp collisions
showing consistency and superior accuracy of the new measurements. It was also
demonstrated that none of the considered models is able to describe all the properties of
\phi production in the measured reactions. Finally, the comparison of \pp to \PbPb
collisions shows a non-trivial system size dependence of the longitudinal
evolution of hidden strangeness production, contrasting with that of other
mesons. The latter observation emphasizes the importance of the  analysis of \BeBe, \ArSc and Xe+La
data collected by \NASixtyOne.
\par
Regarding the search for the QCD critical point, it was shown that there is some
evidence from NA49 for intermittency in \SiSi collisions at \GeVValA{158} and none
detected by NA49 in \CC, \PbPb, nor by \NASixtyOne in \BeBe interactions at the same
energy. The analysis in \ArSc reactions is ongoing.
\par
Spectator-induced electromagnetic effects in pion emission bring
information on the space-time position of the pion formation zone,
which appears to be much closer to the spectator system for faster pions than
for slower ones. On that basis, it has been demonstrated that the longitudinal
evolution of the system at CERN SPS energies can be interpreted as
a pure consequence of local energy-momentum conservation.
\begin{acknowledgement}
This work was supported by the National Science Centre, Poland (grant numbers:
2014/14/E/ST2/00018, 2015/18/M/ST2/00125) and the Foundation for Polish Science
--- MPD program, co-financed by the European Union within the European Regional
Development Fund.
\end{acknowledgement}

\bibliography{bibliography}

\begin{thebibliography}{34}

\bibitem{bib:NA61_facility}
N.~Abgrall et~al. (\NASixtyOne), JINST \textbf{9}, P06005 (2014)

\bibitem{bib:NA49_detector}
S.~Afanasiev et~al. (NA49), Nucl. Instrum. Meth.~A \textbf{430}, 210 (1999)

\bibitem{bib:NA49Onset2002}
S.V. Afanasiev et~al. (NA49), Phys. Rev.~C \textbf{66}, 054902 (2002)

\bibitem{bib:NA49Onset2008}
C.~Alt et~al. (NA49), Phys. Rev.~C \textbf{77}, 024903 (2008)

\bibitem{bib:EPOS2006}
K.~Werner, F.~Liu, T.~Pierog, Phys. Rev.~C \textbf{74}, 044902 (2006)

\bibitem{bib:EPOS2009}
T.~Pierog, K.~Werner, Nucl. Phys. B (Proc. Suppl.) \textbf{196}, 102 (2009)

\bibitem{bib:CRMC}
C.~Baus, T.~Pierog, R.~Ulrich, \emph{{CRMC (Cosmic Ray Monte Carlo package)}}

\bibitem{bib:PythiaManual}
T.~Sjöstrand, S.~Mrenna, P.~Skands, J. High Energy Phys. \textbf{05}, 026
  (2006)

\bibitem{bib:UrQMD1998}
S.~Bass et~al., Prog. Part. Nucl. Phys. \textbf{41}, 255 (1998)

\bibitem{bib:UrQMD1999}
M.~Bleicher et~al., J. Phys.~G \textbf{25}, 1859 (1999)

\bibitem{bib:NA49phi2000}
S.~Afanasiev et~al. (NA49), Phys. Lett.~B \textbf{491}, 59 (2000)

\bibitem{bib:Blobel_pp24_phi_1975}
V.~Blobel et~al., Phys. Lett.~B \textbf{59}, 88 (1975)

\bibitem{bib:ACCMOR_hadrons63_93_phi_1981}
C.~Daum et~al. (ACCMOR), Nucl. Phys.~B \textbf{186}, 205 (1981)

\bibitem{bib:Drijard_ppS53_phi_1981}
D.~Drijard et~al., Z. Phys.~C \textbf{9}, 293 (1981)

\bibitem{bib:AguilarBenitez_pp400_phi_1991}
M.~Aguilar-Benitez et~al. (LEBC-EHS), Z. Phys.~C \textbf{50}, 405 (1991)

\bibitem{bib:Vovchenko2016}
V.~Vovchenko, V.V. Begun, M.I. Gorenstein (2016),
  \texttt{\href{http://arxiv.org/abs/1512.08025}{arXiv:1512.08025 [nucl-th]}}

\bibitem{bib:NA49phi2008}
C.~Alt et~al. (NA49), Phys. Rev.~C \textbf{78}, 044907 (2008)

\bibitem{bib:NA61_pp_spectra_2017}
A.~Aduszkiewicz et~al. (\NASixtyOne), Eur. Phys. J.~C \textbf{77}, 671 (2017)

\bibitem{bib:NA61_pp_pion_2014}
N.~Abgrall et~al. (\NASixtyOne), Eur. Phys. J.~C \textbf{74}, 2794 (2014)

\bibitem{bib:NA49Lambdas2004}
T.~Anticic et~al. (NA49), Phys. Rev. Lett. \textbf{93}, 022302 (2004)

\bibitem{bib:JohnsonRescattering}
S.~Johnson, B.~Jacak, A.~Drees, Eur. Phys. J.~C \textbf{18}, 645 (2001)

\bibitem{bib:Diakonos_CPOD2006}
F.K. Diakonos et~al., PoS \textbf{CPOD2006}, 010 (2006)

\bibitem{bib:Antoniou_2006}
N.G. Antoniou et~al., Phys. Rev. Lett. \textbf{97}, 032002 (2006)

\bibitem{bib:NA49intermittency2015}
T.~Anticic et~al. (NA49), Eur. Phys. J.~C \textbf{75}, 587 (2015)

\bibitem{bib:statReport2017}
A.~Aduszkiewicz et~al. (\NASixtyOne), CERN-SPSC-2017-038; SPSC-SR-221  (2017)

\bibitem{bib:STAR_2014}
L.~Adamczyk et~al. (STAR), Phys. Rev. Lett. \textbf{112}, 162301 (2014)

\bibitem{bib:Rybicki_2015}
A.~Rybicki, A.~Szczurek, M.~Kłusek-Gawenda, Acta Phys. Polon. \textbf{B46},
  737 (2015)

\bibitem{bib:Rybicki_2009}
A.~Rybicki, PoS \textbf{EPS-HEP2009}, 031 (2009)

\bibitem{bib:wa98}
H.~Schlagheck (WA98), Nucl. Phys \textbf{A663}, 725 (2000)

\bibitem{bib:Rybicki_2016}
A.~Rybicki et~al., Acta Phys. Polon. Supp. \textbf{9}, 303 (2016)

\bibitem{bib:Szczurek_2017}
A.~Szczurek, M.~Kiełbowicz, A.~Rybicki, Phys. Rev. \textbf{C95}, 024908 (2017)

\bibitem{bib:Myers_1978}
W.D. Myers, Nucl. Phys. \textbf{A296}, 177 (1978)

\bibitem{bib:trzcinska1}
A.~Trzcińska et~al., Phys. Rev. Lett. \textbf{87}, 082501 (2011)

\bibitem{bib:NA49_2012}
T.~Antitic et~al. (NA49), Phys. Rev. \textbf{C86}, 054903 (2012)

\end{thebibliography}

\end{document}